\documentclass[journal]{IEEEtran}
\usepackage{cite}

\ifCLASSINFOpdf
   \usepackage[pdftex]{graphicx}
  \graphicspath{{figures/}}
   \DeclareGraphicsExtensions{.pdf,.jpeg,.png}
\else
  \usepackage[dvips]{graphicx}
  \graphicspath{{figures/}}
  \DeclareGraphicsExtensions{.eps,.jpeg}
\fi
\usepackage[cmex10]{amsmath}
\ifCLASSOPTIONcompsoc
  \usepackage[caption=false,font=normalsize,labelfont=sf,textfont=sf]{subfig}
\else
  \usepackage[caption=false,font=footnotesize]{subfig}
\fi
\usepackage{url}

\usepackage{amsfonts}
\usepackage{graphicx}
\usepackage{mathtools}
\usepackage{enumerate}
\usepackage{multirow}
\usepackage{todonotes}
\usepackage{array}
\usepackage{tabu}
\usepackage{booktabs}
\usepackage[caption=false,font=footnotesize]{subfig}
\usepackage{bm}

\begin{document}

\title{Robust Acoustic Scene Classification using a Multi-Spectrogram Encoder-Decoder Framework}
%
\author{Lam~Pham,
	Huy~Phan,
	Truc~Nguyen,
	Ramaswamy~Palaniappan, 
	Alfred~Mertins, 
	Ian~McLoughlin%
\thanks{L. Pham, H. Phan and R. Palaniappan are with the University of Kent, School of Computing, Medway, Kent, UK.}%
\thanks{T. Nguyen is with the Signal Processing and Speech Communication Lab, Graz University of Technology, Austria.}%
\thanks{A. Mertins is with the Institute for Signal Processing, University of Lubeck, Germany.}%
\thanks{I. McLoughlin is with the National Engineering Laboratory for Speech and Language Information Processing at the University of Science and Technology of China, Hefei, P.R.China.}%
}

\markboth{IEEE Transactions on Audio, Speech and Language Processing,~Vol.~xx, No.~X, May~20XX}%
{l. Pham\MakeLowercase{\textit{et al.}}: IEEE Transactions on Audio, Speech and Language Processing}

\maketitle
\begin{abstract}
This article proposes an encoder-decoder network model for Acoustic Scene Classification (ASC), the task of identifying the scene of an audio recording from its acoustic signature.
We make use of multiple low-level spectrogram features at the front-end, transformed into higher level features through a well-trained CNN-DNN front-end encoder.
The high level features and their combination (via a trained feature combiner) are then fed into different decoder models comprising random forest regression, DNNs and a mixture of experts, for back-end classification.
%
We report extensive experiments to evaluate the accuracy of this framework for various ASC datasets, including LITIS Rouen and IEEE AASP Challenge on Detection and Classification of Acoustic Scenes and Events (DCASE) 2016 Task 1, 2017 Task 1, 2018 Tasks 1A \& 1B and 2019 Tasks 1A \& 1B. 
%
The experimental results highlight two main contributions; the first is an effective method for high-level feature extraction from multi-spectrogram input via the novel C-DNN architecture encoder network, and the second is the proposed decoder which enables the framework to achieve competitive results on various datasets. The fact that a single framework is highly competitive for several different challenges is an indicator of its robustness for performing general ASC tasks.
\end{abstract}
%
\section{Introduction}
\label{sec:introduction}
%
Considering a general recording of an acoustic environment, this can be said to contain both a background sound field as well as various foreground events.  
If we regard the background as noise and the foreground as signal, we would find that the signal-to-noise ratio exhibits high variability due to the diverse range of environments and recording conditions.
To complicate matters further, a lengthy sound event could be considered background in certain contexts and foreground in others.
For instance, a \textit{pedestrian street} recording may have a generally quiet background, but with short \textit{engine} foreground events, as traffic passes.
However, a lengthy \textit{engine} sound in an \textit{on bus} recording would be considered a background sound.
Furthermore, both background and foreground contain true noise -- continuous, periodic or aperiodic acoustic signals that interfere with the understanding of the scene.
These variabilities and difficulties make acoustic scene classification (ASC) particularly challenging.\\ 
%
To deal with such challenges, recent ASC papers have tended to focus on two main machine hearing areas. 
The first aims to solve the lack of discriminative information by exploiting various methods of trained low-level feature extraction.
In particular, researchers transform audio data into one or two-dimensional frame-based or spectrogram representations to be fed into a back-end classifier.
Frame-based representations often utilise Mel Frequency Cepstral Coefficients (MFCC)~\cite{CUPbook2016}, providing powerful feature extraction capabilities borrowed from the automatic speech recognition (ASR) community~\cite{dca_16_db}. 
MFCCs are often combined with other low-level features such as intensity, zero-crossing rate, etc~\cite{erik_dca_16}, or modified features such as perceptual linear prediction (PLP) coefficients, power nomalised cepstral coefficients (PNCC), robust compressive gamma-chirp filter-bank cepstral coefficients (RCGCC) or subspace projection cepstral coefficients (SPPCC)~\cite{park_dca_16}.
Some systems first transform  audio into spectrograms, then attempt to learn different aspects of those spectrograms to extract frame-based features. 
For instances, Alain et al.~\cite{rako_dca_16} applied non-negative matrix factorisation (NMF) techniques over a Mel-spectrogram. Meanwhile Song et al.~\cite{song_dca_16} applied the auditory statistics of a cochlear filter model to extract discriminative features directly from audio signals. 
Conventionally, frame-based approaches are combined with machine learning methods, such as Gaussian mixture models (GMM)~\cite{dca_16_db, park_dca_16, jung_dca_18}, support vector machines (SVM)~\cite{jung_dca_18, song_dca_16, erik_dca_16} and so on, for the role of back-end classification.
Spectrogram-based approaches tend to use linear or log-Mel stacked short-time fast Fourier transform (STFT) spectra.
Recently published papers show diverse combinations of log-Mel and different types of spectrograms such as Mel-based nearest neighbour filter (NNF) spectrogram~\cite{truc_dca_18, truc_dca_18_icme}, constant-Q transform (CQT)~\cite{hos_dca_18}, or combine two spectrograms, such as MFCC and gammatonegram in~\cite{huy_j01}.
Testing a wavelet-transform derived spectrogram representation, Ren et al.~\cite{dc_17_jou_64} compared results from STFT spectrograms and both \textit{Bump} and \textit{Morse} scalograms.
By exploiting channel information, Sakashita and Aono~\cite{yuma_dca_18} generated multi-spectrogram inputs from two channels, their average and side channels, and even explored separated harmonic and percussive spectrograms from mono channels to achieve good results.
Some papers proposed combining both frame-based and spectrogram features such as i-vector and MFCC spectrogram methods in~\cite{jung_dca_18}, or log-Mel spectrogram and x-vector in~\cite{hos_dca_18}. 
%
Comparing between frame-based and spectrogram representations, the latter provides richer low level feature input detail and appears to enable better performance~\cite{huy_j01,huy_lit_int,huy_lit_aes}.\\
Systems using two-dimensional spectrograms~\cite{ivmDNNsounddet,ivmCNNsounddet} as low-level features tend to be associated with more complex deep learning classifiers. 
In general, input spectrograms are first transformed to high-level features containing condensed information before feeding into a final classifier~
\cite{mcloughlin2017continuous}. 
In some systems, both high-level feature extraction and classification are integrated into one learning process as an end-to-end model.
If the high-level feature transformer is well designed and is effective at obtaining well-represented features, final classifier performance clearly benefits.
From this inspiration, the second research trend focuses on constructing and training powerful learning models to transform spectrograms into well-represented high-level features.
For instance, Lidy and Schindler~\cite{lidy_dca_16} proposed two parallel CNN-based models with different kernel sizes to learn from a CQT spectrogram input, capturing both temporal and frequency information. 
Similarly, Bae et al.~\cite{bae_dca_16} applied a parallel recurrent neural network (RNN) to capture sequential correlation and a CNN to capture local spectro-temporal information over an STFT spectrogram.
Focusing on Pooling layers where high-level features are condensed, Zhao et al.~\cite{zhao_dca_18, zhao_dca_18_ica} proposed an attention pooling layer that showed effective improvement compared to conventional max or mean pooling layers.  
With the inspiration that different frequency bands in a spectrogram contain distinct features, Phaye et al.~\cite{phaye_dca_18} proposed a \textit{SubSpectralNet} network which was able to extract discriminative information from $30$ sub-spectrograms.
Recently, Song et al.~\cite{hong_dca_18} proposed a new way to handle distinct features in a sound scene recording; a deep learning model extracts a bag of features, including both similar and distinct ones, from log-Mel spectrograms, which a back-end network exploits to enhance accuracy.\\
%
Looking at the recent approaches surveyed above, we see three main factors explored by all authors: low-level feature input, high-level feature extraction, and output classification. All of these affect final system accuracy. All are chosen in a task-specific way, and no consensus has emerged regarding an optimum choice for any of the three factors.
In this paper, we address all three factors in the following way:
\begin{enumerate}
\item Firstly, we believe that low level features each contain valuable and complementary information, hence we develop a method to effectively combine three different spectrograms input features, namely log-Mel, Gammatone filter (Gamma) and CQT spectrograms.
\item To extract high-level features from a multi-spectrogram input, we propose a novel encoder-decoder architecture comprising an encoder front end of three parallel CNN-DNN paths (C-DNN).
Each CNN block learns to map one spectrogram into high level features, and we also combine these high level features from the middle layers of the networks to form a combined feature.
\item In terms of decoder as final classifier, we evaluate three different models -- random regression forest, a deep neural network, and a mixture of experts. We compare the performance of each against state-of-the-art approaches.
\end{enumerate}
Rather than selecting a single task, we evaluate over a wide set: LITIS Rouen, DCASE 2016 Task 1, DCASE 2017 Task 1, DCASE 2018 Tasks 1A \& 1B and DCASE 2019 Tasks 1A \& 1B. We will see that the performance of our proposed system is competitive with (and for two tasks, outperforms) the state-of-the-art systems.
The remainder of this paper is organised as follows.
Our motivation for a combined multi-spectrogram approach with a retraining model architecture is described in Section~\ref{sec:system}. 
Section~\ref{sec:eval} describes the evaluation process. Results are discussed in Section~\ref{sec:results}, and we conclude in Section~\ref{sec:conclusion}.
%

\begin{figure}[b]
    \centering
    \includegraphics[width=\linewidth]{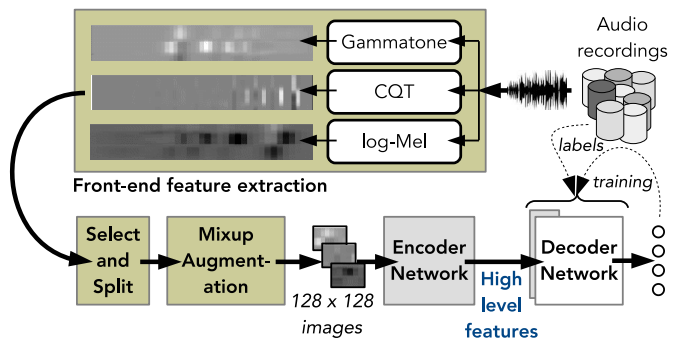}
    	\vspace{-0.7cm}
	\caption{\textit{System view of feature extraction process.}}
    \label{fig:B1}
\end{figure}
\section{The proposed system}
\label{sec:system}
The overall proposed system is outlined in Fig. \ref{fig:B1}.
Firstly, audio from one channel of a recording is represented by a spectrogram.
Having tested numerous spectrogram types in our research, we have found -- and will demonstrate below -- that different spectrograms perform better for different types of scene or task. 
We therefore design an architecture that is able to effectively combine the benefits of log-Mel, Gammatonegram (Gamma) and CQT spectrograms.
The window size, hop size and number of filters is set to 43\,ms, 6 \,ms and 128 for each spectrogram.
Each spectrogram is then split into matching non-overlapping patches of size 128$\times$128.
We apply mixup data augmentation~\cite{mixup1} over the patches to increase  variation, forcing the learning model to enlarge Fisher's criterion.
After mixup, patches are input to the encoder network. 
%
\subsection{Low-level feature with multi-spectrogram input}
\label{ssec:spec}
As mentioned in Section \ref{sec:introduction} and depicted in Fig. \ref{fig:B1}, we employ three different types of spectrograms as low-level features: \\ \\
%
\textbf{a) log-Mel spectrogram (log-Mel)} is popular for ASC tasks, appearing in many recent papers. 
It begins with a set of short-time Fourier transform (STFT) spectra, computed from
\begin{equation}
    \label{eq:stft}
    S(f,t) = \sum_{n=0}^{N-1} \mathbf{x_t}[n]\mathbf{w}[n]e^{-i2{\pi}n{f}/{fs}} 
\end{equation}
where $\mathbf{w}$[$n$] is a window function, typically Hamming, $\mathbf{x_t}$[$n$] is the  discrete audio signal input,  \(N\) is the number of samples per frame, and \(f=k\frac{f_s}{N}\) is the frequency resolution.
A Mel filter bank, which simulates the overall frequency selectivity of the human auditory system using the frequency warping     $f_{mel} = 2595.log(1 + {f}/{700})$~\cite{ian_bk}, is then applied to generate a Mel spectrogram.
Logarithmic scaling is applied to obtain the log-Mel spectrogram.
We use the Librosa toolbox~\cite{librosa_tool} in our experiments. \\ \\
%
\textbf{b) Gammatonegram (Gamma)}: Gammatone filters are designed to model the frequency-selective cochlea activation response of the human inner ear, as given by 
\begin{equation}
    \label{eq:gammaton}
    g(t) = t^{P-1}e^{-2bt\pi}cos(2ft\pi + \theta)    
\end{equation}
where \(t\) is time, \(P\) is the filter order, \(\theta\) is the phase of  the carrier, \(b\) is filter bandwidth, and \(f\) is central frequency.
As with the log-Mel spectrogram, the audio signal is first transformed into STFT spectra before applying a gammatone weighting to obtain the gammatone spectrogram, as in~\cite{auditory2009_tool}. \\ \\
%
\textbf{c) Constant Q Transform (CQT)}: the CQT is designed to model the geometric relationship of pitch, which makes it likely to be effective when undertaking a comparison between natural and artificial sounds, as well as being suitable for frequencies that span several octaves.
As with the log-Mel spectrogram, we also use Librosa~\cite{librosa_tool} to generate the CQT.

Since these spectrograms are derived from different auditory models, it is plausible that they can each contribute distinct features for classification.
This provides an inspiration to explore the three spectrograms. In particular, to design a novel architecture able to extract high-level features from a combination of the three, as described in the following section. 
\subsection{Encoder network for high-level feature extraction}
\label{ssec:pre-train}
\begin{figure}[htb]
    \centering
    \includegraphics[width=0.9\linewidth]{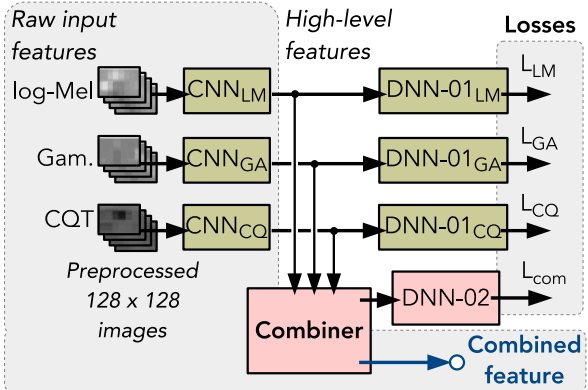}
    	\vspace{-0.1cm}
	\caption{\textit{High-level feature extraction from the encoder network.}}
    \label{fig:B2}
\end{figure}
\begin{table}[htb]
    \caption{Encoder network structures of the CNN (top), DNN-01 (middle) and DNN-02 (bottom).} 
        	\vspace{-0.2cm}
    \centering
    \scalebox{0.9}{

    \begin{tabular}{l c} 
        \hline 
            \textbf{Network architecture}   &  \textbf{Output}  \\
        \hline 
        \textbf{CNN} & \\
         Input layer (image patch) & $128{\times}128$          \\
         Bn - Cv [$9{\times}9$] - Relu - Bn - Ap [$2{\times}2$] - Dr ($0.1\%$)      & $64{\times}64{\times}32$\\
         Bn - Cv [$7{\times}7$] - Relu - Bn - Ap [$2{\times}2$] - Dr ($0.15\%$)      & $32{\times}32{\times}64$\\
         Bn - Cv [$5{\times}5$] - Relu - Bn - Dr ($0.2\%$)      & $32{\times}32{\times}128$ \\
         Bn - Cv [$5{\times}5$] - Relu - Bn - Ap [$2{\times}2$] - Dr ($0.2\%$)       & $16{\times}16{\times}128$\\
         Bn - Cv [$3{\times}3$] - Relu - Bn  - Dr ($0.25\%$)      & $16{\times}16{\times}256$ \\
         Bn - Cv [$3{\times}3$] - Relu - Bn -  Gp - Dr ($0.25\%$) & $256$ \\           
         \hline 
          \textbf{DNN-01} & \\
         Input layer (vector) & $256$ \\
         Fl       &  C         \\
         \hline
         \textbf{DNN-02} &  \\
         Input layer (vector) & $256$ \\
         Fl - Dr ($0.3\%$)        &  $512$       \\
         Fl - Dr ($0.3\%$)        &  $1024$    \\
         Fl       &  C        \\
       \hline 
    \end{tabular}
    }
    \label{table:CDNN} 
\end{table}
High-level features are extracted by the parallel C-DNN front-end paths as shown in Fig. \ref{fig:B2}, referred to as the encoder network.
Image patches of size $128{\times}128$ pixels, after mixup, are fed into the three parallel networks each of which comprises a \textbf{CNN} and a \textbf{DNN-01} block, like the VGG-7 architecture.
The three parallel networks (each containing \textbf{CNN} and  \textbf{DNN-01}) will learn to extract high-level features from one type of spectrogram each.
While the structure of these three CNNs is identical and the structure of the three \textbf{DNN-01} blocks is identical, they will contain very different weights after training.
In Fig.~\ref{fig:B2}, the three paths are denoted by subscripts LM, GA, and CQ, respectively, referring to the kind of spectrogram they process.
The architecture of the \textbf{CNN} and \textbf{DNN-01} blocks is described in the upper and middle sections of Table \ref{table:CDNN}.
Each CNN comprises six layers employing sub-blocks of batch normalization (Bn), convolution (Cv), rectified linear units (Relu), average pooling (Ap), global average pooling (Gp), dropout (Dr), and fully-connected (Fl) layers, with dimensions given in the table.
``C''  is the number of classes found within the given dataset, which depends on the particular evaluation task.

The output of each of the \textbf{CNN} blocks shown in the upper part of Table \ref{table:CDNN} is a 256-dimensional vector.
We refer to the vector extracted from each individual spectrogram, as a high-level feature, and we will explore the relationship between these later.
A size of $256$ was selected after evaluation of power-of-two dimensions from $64$ to $1024$, as well as dimension $128$, on DCASE 2018 Task 1A.

The \textbf{Combiner} block in  Fig.~\ref{fig:B2} has the role of combining the three high level feature vectors into a single combined feature vector.
We will evaluate three methods of combining the high-level features.
Consider vectors $\mathbf{x_{LM/GA/CQ}}$ [$x_{1}, x_{2}, ..., x_{256}$] as the high-level feature outputs of the \textbf{CNN} blocks. 
The first combination method we evaluate, called \textbf{sum-comb}, is  the unweighted sum of the three vectors. i.e. the individual vectors contribute equally to the combined high-level feature,
\begin{equation}
    \label{eq:sum_com}
     \mathbf{x_{sum-comb}} =  \mathbf{x_{LM}} + \mathbf{x_{GA}} + \mathbf{x_{CQ}}
\end{equation}

%
%
The second method, called \textbf{max-comb}, obtains $\mathbf{x_{max-comb}}$ by selecting the element-wise maxim of the three vectors across the dimensions as in eqn. (\ref{eq:max_com}). 
The motivation is to pick the most important (highest magnitude posterior) feature from among the three high level feature vectors,
\begin{equation}
    \label{eq:max_com}
     \mathbf{x_{max-comb}} =  max([ \mathbf{x_{LM}, x_{GA}, x_{CQ}]}, axis=0)
\end{equation}

For the final method, we assume elements of three vectors to have a linear relationship across dimensions. We then derive a simple data-driven combination method called \textbf{lin-comb} by employing a fully-connected layer trained to weight and combine the three high level features, as in

\begin{eqnarray}
    \label{eq:lin_com}
     \mathbf{x_{lin-comb}} =  ReLu \left\{ \mathbf{x_{LM}.w_{LM}} + \right. ...\nonumber \\
\left.      \mathbf{x_{GA}.w_{GA} + x_{CQ}.w_{CQ} + w_{bias}}  \right\}
\end{eqnarray}
where $\mathbf{w_{LM/GA/CQ/bias}}$[$w_{1}, w_{2}, ..., w_{256}$] are trained parameters of the same dimension as the high-level feature vectors.  

The combined high-level feature vector from the output of the \textbf{Combiner} block is then fed into \textbf{DNN-02}, with the structure shown in the lower part of Table \ref{table:CDNN}. 
Note that the combined high-level feature vectors, like the individual high level vectors, have a dimension of 256 -- meaning that the higher layer classifier of the decoder can be set for evaluation with either individual or combined feature input, without changing its structure or the number of trainable parameters it has.

Regarding training loss, we define four loss functions to train the encoder network; three to optimize individual spectrograms, and the final one for their combination.
Eventually, the overall loss function is computed from
\begin{equation}
    \label{eq:loss_func}
    Loss_{EN} =  \alpha(L_{LM} + L_{GA} + L_{CQ}) + \beta.L_{com}
\end{equation}
where \(Loss_{EN}\) is the overall loss function of the encoder network extracting high-level features from three spectrograms.
\(L_{LM}, L_{GA}, L_{CQ}$ and $L_{com}\) are individual losses from the log-Mel, Gamma and CQT spectrograms, and their combinations. These are depicted in Fig.~\ref{fig:B2} and will be defined in Section~\ref{ssec:setup}.
The balancing parameters \(\alpha\) and \(\beta\) focus on learning particular features or combinations and are set to $1/3$ and $1.0$ here, making the contributions from each spectrogram equal.

%
\subsection{Decoders for back-end classification}
\label{ssec:post-train}
\begin{figure}[tb]
    \centering
    \includegraphics[width=1.0\linewidth]{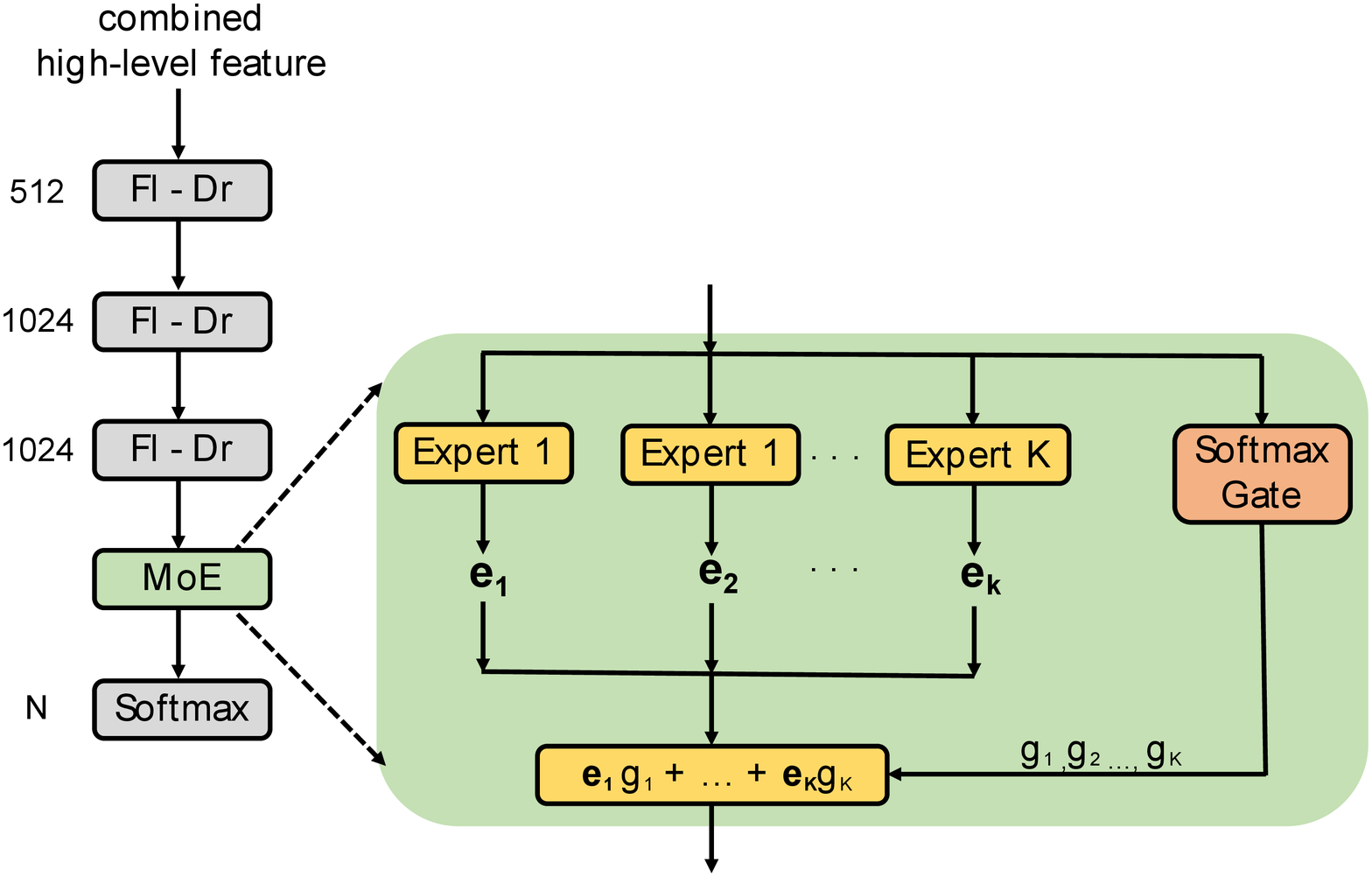}
    	\vspace{-0.2cm}
	\caption{\textit{Proposed mixture of experts within its deep back-end decoder network.}}
    \label{fig:C1}
        	\vspace{-2mm}
\end{figure}

Our previous work~\cite{lam_dca_18}, which introduced a C-DNN structure for individual spectrograms, found mixup to be beneficial for training feature extractors.
The new architecture introduces a feature combiner, so we maintain the previous mixup for low-level features to help train the encoder, but additionally include a second mixup stage, for high-level features when training the decoder.
Furthermore, we will evaluate three types of decoder: A  random forest regressor (RFR) with classifier, a DNN, and a mixture of experts (MoE), described below \\ \\
%
\textbf{a) Random Forest Regression (RFR-decoder)}: 
A regression forest is a type of ensemble model, comprising multiple regression trees. 
The role of each tree is to map the complex input space defined by the multidimensional high-level features from the encoder network, into a continuous class-dimension output space. 
Its nonlinear mapping is achieved by dividing the large original input space into smaller sub-distributions. 
Individual trees are trained using a subset randomly drawn from the original training set. 
By using many trees ($100$ in total), the structure is effective at tackling overfitting issues that can occur with single trees.
We also believe the regressor structure benefits from the continuous mixed-class training labels provided by the use of mixup.
Eventually, the decoded output spaces are classified as in our previous work~\cite{huy_dca_16, huy_j01} by mean pooling the output over all trees.  \\ \\
\textbf{b) Deep Neural Network (DNN-decoder)}:
In this paper, we propose a DNN architecture, denoted  \textbf{DNN-03} for output classification in the decoder.
The network comprises four fully connected dense blocks with dropout, having node sizes of $512-1024-1024-C$, where ``C'' is the number of classes in the task being evaluated.
Note that this is similar to the \textbf{DNN-02} architecture in Fig. \ref{fig:B1}, 
but incorporates one additional fully-connected and one dropout layer, which is useful in practice to refine the accuracy for similar classes.  \\ \\
%
\textbf{c) Mixture of Experts (MoE-decoder)}:
An MoE is a machine learning technique that divides the problem spaces into homogeneous regions by using an array of different trained (but in this case identical structure) models, referred to as experts~\cite{lung_moe}.
A conventional MoE architecture comprises many experts and incorporates a gate network to decide which expert is applied in which input region.
In this paper, we apply the MoE technique to the combined high-level features, as shown in Fig. \ref{fig:C1}.
Specifically, the 256-dimensional input vector goes through three dense layers with dropout, having 512, 1024 and 1024 hidden nodes, respectively, matching \textbf{DNN-01} and \textbf{DNN-02} in the number of hidden units.
The output enters the MoE layer, which is expanded in Fig. \ref{fig:C1}.
The combined result from the experts is gated before passing through a softmax to determine the final $C$ class scores.
Each MoE expert comprises a dense block with a ReLu activation function. Its input dimension is 256 and its output size is $C$.
The gate network is implemented as a \textbf{Softmax Gate} --  an additional fully-connected layer with softmax activation function and a gating dimension equal to the number of experts.

Let  $\mathbf{e_{1}, e_{2}}, \dots  \mathbf{e_{K}} \in \mathbf{R^{C}}$ be the output vectors of the $K$ experts, and  $g_{1}, g_{2}, \dots , g_{K}$ be the outputs of the gate network where $g_k \in \mathbb{R}, \sum_{k=1}^K{g_k}=1$
The predicted output is then found as,
\begin{equation}
    \label{eq:moe}
    \hat{y} = softmax \left\{ \sum_{k=1}^{K} \mathbf{e_{k}}g_{k} \right \}.
\end{equation}
%
%
\section{Evaluation methodology}
\label{sec:eval}
To clearly demonstrate the general performance of the proposed systems we will evaluate using five different ASC tasks.
While it is relatively easy to perform well in one challenge, it is considerably more difficult to do so for all -- this helps us to explore one of the hypothesised strengths of our combined-spectrogram approach, that it can be more generic.
Four of the datasets we use are derived from annual DCASE challenges, whereas the fifth is the extensive LITIS Rouen dataset. Each is described below.

\subsection{DCASE datasets}
We adopt datasets from the DCASE 2016, 2017, 2018 and 2019 challenges.
For DCASE 2016, both development set (1170 segments) and evaluation set (390 segments) were published. Recordings were sampled at 44100\,Hz, with 30\,s duration per segment over the 15-class challenge.
DCASE 2017 reused all DCASE 2016 dataset and added more recording data, with 4680 and 1620 segments for development and evaluation, respectively (segments are of 10\,s duration).
For DCASE 2016 and 2017, we use the development set (dev. set) for training, and report the classification accuracy over the evaluation set (eva. set).
DCASE 2018 and 2019 Task 1A datasets contain 10\,s segments, recorded at 48000\,Hz and spanning 10 classes.
Unlike DCASE 2016 and 2017, these recent challenges only released the development set publicly, providing 8640 and 13370 segments for DCASE 2018 and 2019, respectively.
Moreover, DCASE 2018 and 2019 also proposed a different ASC challenge type that involves mismatched recording devices. This is known as Task 1B. 
Specifically, all recorded segments from device A for the conventional ASC task (the 1A dataset) were reused in Task 1B. 
Then additional segments were recorded using two different devices (B \& C), and added, but with unbalanced recording times of 4 and 6 hours respectively. This is much less than the approximately 24 and 37 hours of device A recordings included in DCASE 2018 and 2019, respectively.  
In this paper, we follow the setting of the DCASE 2018 challenge; subdividing the development dataset into two subsets; a training set (train set) and a testing set (test set), respectively.
For the most recent DCASE 2019 dataset, we report the results over the eva. set (noting that this set has not been released publicly yet), as used for the Kaggle competition associated with the DCASE 2019 challenge\footnote{1A: \url{https://www.kaggle.com/c/dcase2019-task1a-leaderboard/overview}\\1B: 
\url{https://www.kaggle.com/c/dcase2019-task1b-leaderboard/leaderboard}}.

\subsection{LITIS Rouen dataset}
This extensive dataset comprises 19 urban scene classes with 3026 segments, divided into 20 training/testing splits. 
The audio was recorded at a sample rate of 22050\,Hz, with each segment duration being 30\,s.  
We follow the mandated settings for 20 times cross validation, obtaining final classification accuracy by averaging over the 20 testing folds.

\subsection{Experimental setup}
\label{ssec:setup}
All of our proposed networks are built on the Tensorflow framework using cross-entropy loss,
\begin{equation}
    \label{eq:loss_func}
    Loss(\theta) = -\frac{1}{N}\sum_{i=1}^{N}\mathbf{y_i} .log \left\{\mathbf{\hat{y}_{i}}(\theta) \right\} + \frac{\lambda}{2}.||\theta||_{2}^{2}
\end{equation}
where \(Loss(\theta)\) is the loss function over all parameters \(\theta\), constant \(\lambda\) is set initially to $0.0001$,  $\mathbf{y_{i}}$ and $\mathbf{\hat{y}_{i}}$  are expected and predicted results, respectively.
Experiments use the Adam optimiser to adjust learning rate, with a batch size of $50$. 
Results were obtained after $200$ epochs (in practice we lose only a small degree of performance by not continuing beyond this).
As aforementioned, we also performed mixup data augmentation.
For the pre-training process on the extractor, each of the raw $128{\times}128$ dimensional feature was repeated twice by including same-dimension beta and Gaussian distribution mixup images of the same dimensions.
When training the decoder, we applied mixup to the high-level feature vectors prior to the final classifier. 
In this case it doubles the number of $256$ element feature vectors by including same-length beta distribution mixup vectors.
In each case, we incorporated both original and generated mixup data into the training processes to improve performance, at the cost of considerably increasing the training time.

\section{Experimental results and comparison}
\label{sec:results}
In this section we will analyse the performance of the encoder network to specifically understand the contribution made by different spectrogram types, as well as their combinations. 
We will then analyse the performance of the decoder to assess different back-end classifiers, then compare overall performance to a range of state-of-the art methods.

\subsection{The performance of each spectrogram by class}
\label{ssec:Compare_spec}
We first evaluated the baseline architecture to determine how different spectrogram types contributed to the performance of different classes.
To do this, we began by training three C-DNN encoder networks, comprising \textbf{CNN} and \textbf{DNN-02} blocks, each encoder for an individual spectrogram input. 
We trained another C-DNN encoder network, the entire network as in Fig. \ref{fig:B2}, for spectrogram combination.
These four trained systems were subsequently used as high-level feature extractors to train the decoder and then to test the overall system. 
We trained four different decoders using the \textbf{DNN-03} architecture from Section~\ref{ssec:post-train} to assess individual spectrogram performance, as well as the performance of the combined high level features -- using the \textbf{lin-comb} method from Section~\ref{ssec:pre-train}.
These experiments were conducted using the DCASE 2018 Task 1B dev. set.

\begin{figure}[tb!]
    \centering
    \includegraphics[width=0.95\linewidth]{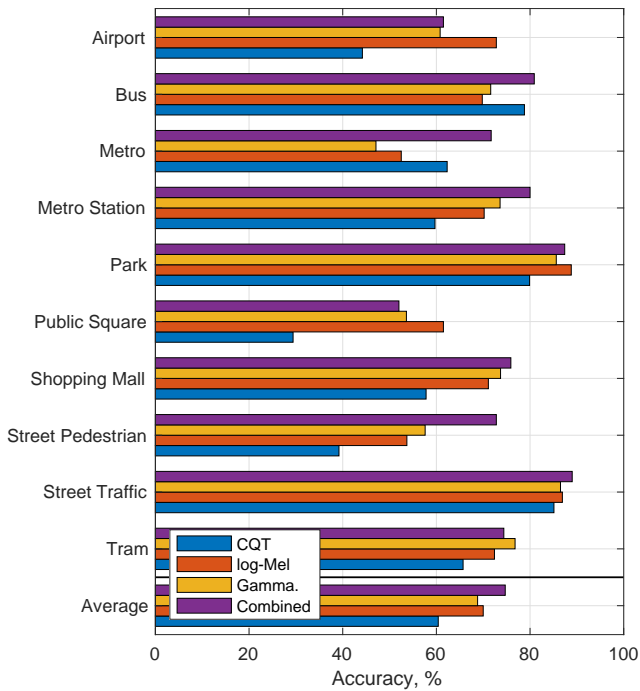}
    	\vspace{-0.1cm}
	\caption{\textit{Performance comparison of different spectrograms types, and their combination, for the DCASE 2018 Task 1B dev. set.}}
	    	\vspace{-0.2cm}
    \label{fig:Z4}
\end{figure}

To compare performances, class-wise accuracies for the three spectrograms and their combinations is shown in Fig. \ref{fig:Z4}, with overall average performance shown at the bottom.
Clearly, the combined features performed best overall, with the log-Mel and gammatonegram performing similarly, and both being better than CQT.
However a glance at the per-class accuracy shows some interesting variation. For example, the CQT spectrogram was particularly good at discriminating the \textit{Bus} and \textit{Metro} classes, compared to the other spectrograms.
Also, while log-Mel and Gamma performances were similar, the former excelled on \textit{Airport} and \textit{Public Square} classes, whereas the latter tended to be slightly better for classes containing vehicular sounds (with the exception of the \textit{Metro} class).

We conclude from this that the three spectrograms represent sounds in  ways that have affinity for certain types of sounds (mirroring a conclusion in~\cite{ivm_cssp2019}, albeit on very different types of sound data). It is therefore unsurprising that intelligently combining the three spectrograms into a high level feature vector can achieve significant performance gain over single spectrograms. 
\subsection{Spectrogram performance for each device}
%
\begin{figure}[bt]
    \centering
    \includegraphics[width=0.8\linewidth]{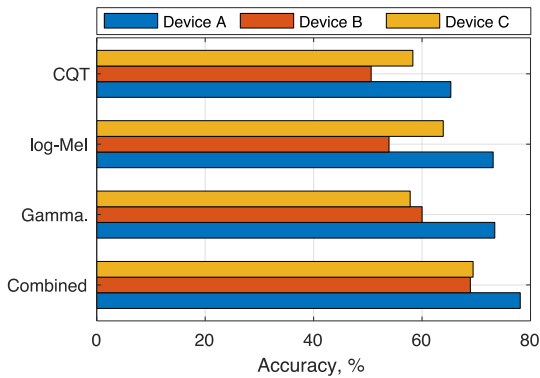}
    	\vspace{-0.2cm}
	\caption{\textit{Performance comparison for different recording devices within the DCASE 2018 Task 1B dev. set.}}
    \label{fig:Z3}
\end{figure}
%
%
\begin{figure}[hbt]
    \centering
    \includegraphics[width=0.9\linewidth]{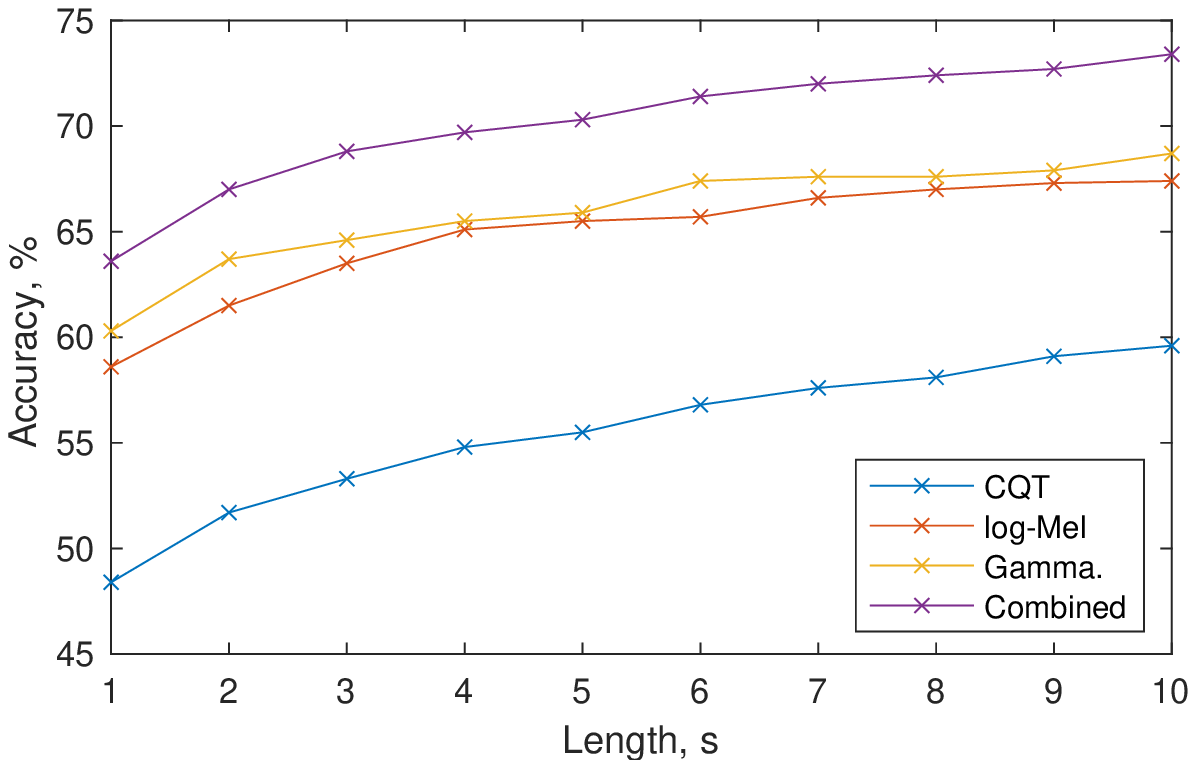}
    	\vspace{-0.2cm}
	\caption{\textit{Classification performance as a function of the length of the test signal over DCASE 2018 Task 1B dev. set - all devices.}}
    \label{fig:Z5}
\end{figure}
%
\begin{figure}[hbt]
    \centering
    \includegraphics[width=0.9\linewidth]{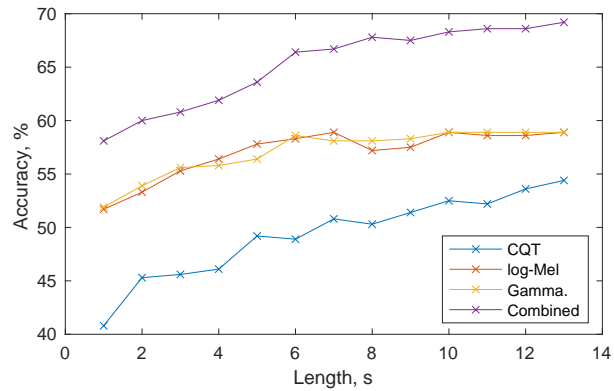}
    	\vspace{-0.2cm}
	\caption{\textit{Classification performance as a function of the length of the test signal over DCASE 2018 Task 1B dev. set - devices B\&C.}}
    \label{fig:Z6}
\end{figure}
DCASE 2018 task 1B  includes highly unbalanced data recordings from three different devices as described in Section~\ref{sec:eval}.
We analysed the performance of different spectrograms for those three devices, with results plotted in Fig. \ref{fig:Z3}.
The device with the largest amount of training data (Device A) obviously scored best, achieving around 9\% better accuracy than devices B and C.
Again, the Gamma and log-Mel results were similar, but each `preferred' a different minority device. 
Although there were not enough devices included in the test for the evidence to be conclusive, this variability indicates that spectrograms differ in their affinity for different devices (or device locations, or channels). Again, the combined features effectively leveraged the advantages of each spectrogram type.

\subsection{Spectrogram performance by segment length}
\begin{table*}[bt]
    \caption{Comparison to DCASE 2018 baselines for Task 1B dev. set (using \textbf{lin-comb} for the pre-training process).} 
   	\vspace{0.3cm}
    \centering
    \begin{tabular}{l | c c c c  |  c c c c}
       \hline
& \multicolumn{4}{l}{\textbf{Device A}} & \multicolumn{4}{l}{\textbf{Devices  B \& C }} \\ 
Classes				&D.2018      &RFR-decoder      &DNN-decoder      &MoE-decoder    &D.2018       &RFR-decoder         &DNN-decoder      &MoE-decoder       \\                     
	 \hline                                                            
             Airport            &73.4        &67.5     &60.4     &66.8    &72.5        &55.6        &69.4     &75.0      \\
             Bus                &56.7        &78.5     &80.2     &80.2   &78.3        &88.9        &86.1     &88.9      \\
             Metro              &46.6        &67.0     &72.8     &69.3   &20.6        &75.0        &63.9     &66.7      \\
             Metro Stn.      &52.9        &84.6     &82.6     &80.3   &32.8        &50.0        &61.1     &50.0      \\
             Park               &80.8        &89.7     &86.8     &88.4   &59.2        &91.7        &91.7     &94.4      \\
             Pub. Sq.      &37.9        &47.7     &52.8     &50.9   &24.7        &52.8        &47.2     &47.2      \\
             Shop. Mall      &46.4        &74.6     &75.3     &73.8   &61.1        &80.6        &80.6     &80.6      \\
             Str. Ped.  &55.5        &65.6     &72.5     &71.3   &20.8        &66.7        &75.0     &77.8      \\
             Str. Traffic     &82.5        &91.1     &90.7     &92.3   &66.4        &75.0        &77.8     &77.8      \\
             Tram               &56.5        &83.1     &79.3     &83.1   &19.7        &52.8        &38.9     &47.2      \\
       \hline                                                                             
	     Average    &\textbf{58.9}  &\textbf{75.2} &\textbf{75.5} &\textbf{75.9}  &\textbf{45.6}     &\textbf{68.9}     &\textbf{69.2}     & \textbf{70.6}    \\   
       \hline 
    \end{tabular}    
    \label{table:comp_dcase} 
\end{table*}
\begin{table}[bt]
    \caption{Performance of re-trained models (encoder/decoder Acc. \%) over DCASE 2018 Task 1B dev. set.} 
        	\vspace{0.2cm}
    \centering
{
    \begin{tabular}{l | c c c}
\textbf{Device A }   &RFR-decoder     &DNN-decoder      &MoE-decoder    \\                     
        \hline 
        sum-comb  &71.5/75.6              &71.5/72.2               &71.5/71.9    \\
        max-comb   &74.1/75.3             &74.1/74.7                &74.1/75.5     \\
        lin-comb     &73.7/75.2             &73.7/75.5                &73.7/\textbf{75.9}  \\
\hline 
\hline 
\textbf{Devices  B \& C:}    &RFR-decoder         &DNN-decoder      &MoE-decoder       \\                     
\hline 
        sum-comb    &63.9/64.4            &63.9/65.6     &63.9/63.9   \\
        max-comb  &61.4/65.3            &61.4/63.9     &61.4/63.9   \\
        lin-comb     &64.2/68.9    &64.2/69.2      &64.2/\textbf{70.6}   \\
    \end{tabular}
    }
    \label{table:re_model} 
\end{table}
%
Inspired by a number of previous works that considered the ability of systems to recognise a sound class early~\cite{huyearly_2018,ivmearly_2018}, we also evaluated this ability for the different spectrogram types.
Figs. \ref{fig:Z5} and \ref{fig:Z6} plot early classification accuracy for DCASE 2018 Task 1B for all devices and for devices B$+$C, respectively. Early classification means that class assignment is performed only on the first part of the audio recording, rather than the entire duration (i.e. on cropped audio). Performance is plotted for a number of cropped segment lengths between 1\,s and the full 10\,s.

From both plots, immediate observations are that the combined high level features performed much better than the individual spectrogram types. 
The CQT performed worst while the other two spectrograms had similar performance (as in the experiments above).
Looking closer at Fig.~\ref{fig:Z5} (accuracy for all devices), the score for all features continued to climb as duration progressed towards the full 10\,s. 
This provides a strong indication that the system was data-constrained and is likely to perform better with longer duration recordings.
By contrast, Fig.~\ref{fig:Z6} contains indications that the performance of the log-Mel and Gamma spectrograms began to plateau as duration exceeded 5\,s, indicating that performance might not substantially increase if longer duration recordings were available. 
However the continued improvement of the CQT representation as  length increased gave the combined features an ability to gain higher accuracy from longer recordings: The strength of CQT may lie in the analysis of longer recordings.

However, in these experiments, CQT performance lagged the combined features by around 15\% absolute, with the other spectrograms lagging by only around 5\% absolute -- apart from the area in Fig.~\ref{fig:Z6} where they plateaued.

Most remarkable, though, is that with just 2\,s of input data from a recording, our proposed combined high level feature was able to match or outperform any of the individual spectrograms operating with the full 10\,s of input data.
This clearly demonstrates a major advantage of the proposed system. 
It effectively captures the advantages of the individual spectrogram features, which vary in their affinity for different classes and devices, and yields extremely good performance even when a restricted amount of data is available for classification.
\subsection{Performance of different classifiers in the decoder}
\label{ssec:compare_postrain}
\begin{table*}[!bt]
    \caption{Comparison of the proposed system (\textbf{lin-comb}+\textbf{MoE-decoder}) to state-of-the-art results, with best performance in \textbf{bold} (Upper part: Dataset; Middle part: top-ten DCASE challenges; Lower part: State-of-the-art papers) } 
    \centering
    \vspace{0.3cm}
    \resizebox{\textwidth}{!}{%
    \begin{tabular}{l c  |  l c  | l c  | l c |  l c | l c || l c} 
\textbf{D.2016}              &\textbf{Acc.}    &\textbf{D.2017}              &\textbf{Acc.}   &\textbf{D.2018-1A}              &\textbf{Acc.}        &\textbf{D.2018-1B}                  &\textbf{Acc.}     &\textbf{D.2019-1A}         &\textbf{Acc.}       &\textbf{D.2019-1B}                       &\textbf{Acc.}     &\textbf{LITIS}                &\textbf{Acc}\\ [0.5ex] 
                                                                                                                                                                                                                                                                                                                                                               
\multicolumn{2}{l|}{ \textbf{(eva. set)}}     & \multicolumn{2}{l|}{ \textbf{(eva. set)}}    & \multicolumn{2}{l|}{ \textbf{(dev. set)}}             &\multicolumn{2}{l|}{ \textbf{(dev. set)}}              &\multicolumn{2}{l|}{ \textbf{(eva. set)}}       &\multicolumn{2}{l|}{ \textbf{(eva. set)}}                   &\multicolumn{2}{l}{\textbf{(20-fold ave.)} }    \\
                                                                                                                                                                                                                                                                                                                                                               
\hline                                                                                                                                                                                                                                                                                                                                            
Wei~\cite{dc_2016_t10}        &$84.1$          &Zhao~\cite{dc_2017_t10}      &$70.0$          &Li~\cite{dc_2018_t10}           &$72.9$               &Baseline~\cite{dc_2018_bsl}         &$45.6$            &Mingle~\cite{dc_2019_t10}  &$79.9$              &Baseline~\cite{dc_2019_dataset}          &$61.6$            &Bisot~\cite{bisot2015hog}     &$93.4$  \\     
Bae~\cite{bae_dca_16}         &$84.1$          &Jung~\cite{dc_2017_t09}      &$70.6$          &Jung~\cite{jung_dca_18}         &$73.5$               &Li~\cite{dc_2018_tb07}              &$51.7$            &Wu~\cite{dc_2019_t09}      &$80.1$              &Kong~\cite{dc_2019_tb09}                 &$61.6$            &Ye~\cite{ye_lit_02}           &$96.0$  \\     
Kim~\cite{dc_2016_t08}        &$85.4$          &Karol~\cite{dc_2017_t08}     &$70.6$          &Hao~\cite{dc_2018_t08}          &$73.6$               &Tchorz~\cite{dc_2018_tb06}          &$53.9$            &Gao~\cite{dc_2019_t08}     &$80.5$              &Waldekar~\cite{dc_2019_tb08}             &$62.1$            &Huy~\cite{huy_dca_16}         &$96.4$  \\     
Takahasi~\cite{dc_2016_t07}   &$85.6$          &Ivan~\cite{dc_2017_t07}      &$71.7$          &Christian~\cite{dc_2018_t07}    &$74.7$               &Kong~\cite{dc_2018_tb05}            &$57.5$            &Wang~\cite{dc_2019_t07}    &$80.6$              &Wang~\cite{dc_2019_tb07}                 &$70.3$            &Yin~\cite{yin2018learning}    &$96.4$  \\     
Elizalde~\cite{dc_2016_t06}   &$85.9$          &Park~\cite{dc_2017_t06}      &$72.6$          &Zhang~\cite{dc_2018_t06}        &$75.3$               &Wang~\cite{dc_2018_tb04}            &$57.5$            &Jung~\cite{dc_2019_t06}    &$81.2$              &Jiang~\cite{dc_2019_tb06}                &$70.3$            &Huy~\cite{huy_j01}            &$96.6$  \\     
Valenti~\cite{dc_2016_t05}    &$86.2$          &Lehner~\cite{dc_2017_t05}    &$73.8$          &Li~\cite{dc_2018_t05}           &$76.6$               &Waldekar~\cite{walde}               &$57.8$            &Huang~\cite{dc_2019_t05}   &$81.3$              &Song~\cite{dc_2019_tb05}                 &$72.2$            &Ye~\cite{ye_lit}              &$97.1$  \\     
Marchi~\cite{erik_dca_16}     &$86.4$          &Hyder~\cite{dc_2017_t04}     &$74.1$          &Dang~\cite{dc_2018_t04}         &$76.7$               &Zhao~\cite{zhao_dca_18}             &$58.3$            &Haocong~\cite{dc_2019_t04} &$81.6$              &Primus~\cite{dc_2019_tb04}               &$74.2$            &Huy~\cite{huy_lit_int02}      &$97.8$  \\     
Park~\cite{park_dca_16}       &$87.2$          &Zhengh~\cite{dc_2017_t03}    &$77.7$          &Octave~\cite{octave_exploring}  &$78.4$               &Truc~\cite{truc_dca_18}             &$63.6$            &Hyeji~\cite{dc_2019_t03}   &$82.5$              &Hamid~\cite{dc_2019_tb03}                &$74.5$            &Zhang~\cite{zhang_lit_int02}  &$97.9$  \\  
Bisot~\cite{Bisot2016}        &$87.7$          &Han~\cite{dc_2017_t02}       &$80.4$          &Yang~\cite{yang_acoustic}       &$79.8$               &                                    &                  &Koutini~\cite{dc_2019_t02} &$83.8$              &Gao~\cite{dc_2019_tb02}                  &$74.9$            &Zhang~\cite{zhang_lit_int}    &$98.1$  \\     
Hamid~\cite{Eghbal-Zadeh2016} &$89.7$          &Mun~\cite{dc_2017_t01}       &$\textbf{83.3}$ &Golubkov~\cite{dc_2018_t01}     &$\textbf{80.1}$      &                                    &                  &Chen~\cite{dc_2019_t01}    &$\textbf{85.2}$     &Kosmider~\cite{dc_2019_tb01}             &$\textbf{75.3}$   &Huy~\cite{huy_lit_int}        &$98.7$  \\ 
                                                                                                                                                                                                                                                                                                                                   
\cmidrule{1-8}                                                                                                                                                                                                                                                                                                            
Mun~\cite{mun2017ica}         &$86.3$          &Zhao~\cite{dc_17_jou_64}      &$64.0$         &Bai~\cite{bai}                  &$66.1$               &Zhao~\cite{zhao_dca_18_ica}         &$63.3$            &                           &                    &                                         &                  & & \\   
Li~\cite{li2017icas}          &$88.1$          &Yang~\cite{dc_17_ica_693}     &$69.3$         &Gao~\cite{gao_dc18}             &$69.6$               &Truc~\cite{truc_dca_18_int}         &$64.7$            &                           &                    &                                         &                  & & \\   
Hyder~\cite{hyder2017int}     &$88.5$          &Waldekar~\cite{dc_17_int_699} &$69.9$         &Zhao~\cite{zhao_dca_18_ica}     &$72.6$               &Truc~\cite{truc_dca_18_icme}        &$66.1$            &                           &                    &                                         &                  & & \\   
Song~\cite{Song2018int}       &$89.5$          &Wu~\cite{dc_17_ica_754}       &$75.4$         &Phaye~\cite{phaye_dca_18}       &$74.1$               &                                    &                  &                           &                    &                                         &                  & & \\   
Yin~\cite{yin2018learning}    &$\textbf{91.0}$ &Chen~\cite{dc_17_ica_771}     &$77.1$         &Heo~\cite{heo}                  &$77.4$               &                                    &                  &                           &                    &                                         &                  & & \\    
                                                                                                                                                                                                                                                                                                                                                                  
\hline				                                                                                                                                                                                                                                                                                                                          
Our system                    &$88.2$          &Our system                    &$72.6$          &Our system                     &$77.5$               &Our system                          &$\textbf{70.6}$   &Our system                 &$76.8$              &Our system                               &$72.8$            &Our system                    &$\textbf{98.9}$    \\
    \end{tabular} }   
    \label{table:state_of_the_art} 
\end{table*}
%
Three methods were proposed in Section~\ref{ssec:pre-train} to incorporate the three high level spectrogram features into a combined high-level feature in the encoder network.
These were namely \textbf{sum-comb, max-comb} and \textbf{lin-comb}. 
To make use of the combined features, we then introduced three back-end classifier methods for the decoder block, namely \textbf{RFR-decoder, DNN-decoder} and \textbf{MoE-decoder} in Section~\ref{ssec:post-train}.

In total, the three classifiers and three combiners yield 9 models to evaluate.
In this section, we compare performance among these 9 models on the DCASE 2018 Task 1B dev. dataset.
We separately note the accuracy of the encoder network (i.e. the feature extractor, alone), as well as the full system accuracy (i.e. incorporating the decoder).

Results are presented in Table \ref{table:re_model}, again split into Device A and Device B \& C performance.
Best performance for both device sets, highlighted in bold, was achieved by the \textbf{MoE-decoder} classifier with the \textbf{lin-comb} combiner. 
However some interesting trends were evident. Firstly, \textbf{DNN-decoder} was only very slightly inferior to \textbf{MoE-decoder} for all combiners and device types.
Secondly, looking at the encoder network results for the Device A evaluation, the  \textbf{max-comb} combiner actually outperformed the accuracy of \textbf{lin-comb}, although the latter performed best for most of the full systems. 
This means that the optimal high level feature combiner for the full system was not the best combiner for loss computation when training the encoder network.
However the situation reverses when looking at Devices B \& C -- an indication that the performance gain of \textbf{lin-comb} may have been due to better generalisation.

%
\subsection{Per-class performance of different decoders}
\label{ssec:compare_dcase}
Given that the results presented so far indicate that the \textbf{lin-comb} combiner performed best, we now feed those high level combined features into the three alternative decoders to explore class-by-class performance.
Table \ref{table:comp_dcase} presents results for DCASE 2018 Task 1B (dev. dataset). Device A and Device B \& C results are again shown separately, and the ``D.2018'' column is the DCASE 2018 baseline.
Results show that the three classifiers all outperformed the baseline -- with the mixture of experts system improving accuracy by 16.6\% and 24.0\% absolute, for Device A and Devices B \& C, respectively.

%
\subsection{Performance comparison to state-of-the-art systems}
\label{ssec:state_of_the_art}
While performance against the baseline score of DCASE 2018 is good, 
we now evaluate the same model configuration (i.e. \textbf{lin-comb} combiner and \textbf{MoE-decoder} back-end classifier) on various datasets and competitions, to compare performance against the state of the art at time of writing.
The results, listed in Table \ref{table:state_of_the_art}, show that the system proposed in this paper achieves the highest accuracy for two datasets -- achieving 70.6\% and 98.9\%for DCASE 2018 Task 1B dev. and LITIS Rouen, respectively.
For DCASE 2016, an accuracy of 88.2\% was achieved, taking second position on the challenge table, and ranked top-four among state-of-the-art systems. DCASE 2017 performance is a little less competitive at 72.6\% (note that the system used for that was slightly modified in that it normalized the input data).
Our DCASE 2018 Task 1A performance was 77.5\%, taking third place on the challenge table.
We also entered the system to the recent DCASE 2019 challenge, achieving  76.8\% and 72.8\% for DCASE 2019 Task 1A and 1B, respectively. 
(Noting that recorded accuracy in Table \ref{table:state_of_the_art} are collected from published papers and technique reports)

\section{Conclusion}
\label{sec:conclusion}
This paper has presented a robust framework for acoustic scene classification. 
Using a feature approach based upon three kinds of time-frequency transformation (namely log-Mel, gammatone filter and constant Q transform), we presented a two-step training method to first train a front-end encoder network, and then train a decoder to perform back-end classification.
To deal with the many challenges implicit in the ASC task, we investigated how the different time-frequency spectrogram types can be combined effectively to improve classification accuracy. 
In terms of results, the classification accuracy obtained from the proposed system, comprising a trained feature combiner and utilising an MoE-based decoder performs particularly well. 
Evaluated on DCASE and LITIS Rouen datasets, the proposed method achieves highly competitive results compared to state-of-the-art systems for all tasks, in particular achieving the highest LITIS Rouen and DCASE 2018 Task 1B accuracy at the time of writing.

\vfill\pagebreak


\bibliographystyle{IEEEbib}
\bibliography{strings,refs}

\end{document}